# SWENet: a physics-informed deep neural network (PINN) for shear wave elastography

Ziying Yin[1, †], Guo-Yang Li[2, †], Zhaoyi Zhang[1], Yang Zheng[1], Yanping Cao[1, *]

[1] Institute of Biomechanics and Medical Engineering, AML, Department of Engineering Mechanics, Tsinghua University, Beijing 100084, China.

[2] Harvard Medical School and Wellman Center for Photomedicine, Massachusetts General Hospital, Boston, Massachusetts 02114, USA.

[†] Co-first authors with equal contribution.

[*] Corresponding author: caoyanping@tsinghua.edu.cn (Y.C.).

**ABSTRACT**

Shear wave elastography (SWE) enables the measurement of elastic properties of soft materials, including soft tissues, in a non-invasive manner and finds broad applications in a variety of disciplines. The state-of-the-art SWE methods commercialized in various instruments rely on the measurement of shear wave velocities to infer material parameters and have relatively low resolution and accuracy for inhomogeneous soft materials due to the complexity of wave fields. In the present study, we overcome this challenge by proposing a physics-informed neural network (PINN)-based SWE (SWENet) method considering the merits of PINN in solving an inverse problem. The spatial variation of elastic properties of inhomogeneous materials has been defined in governing equations, which are encoded in PINN as loss functions. Snapshots of wave motion inside a local region have been used to train the neural networks, and during this course, the spatial distribution of elastic properties is inferred simultaneously. Both finite element simulations and tissue-mimicking phantom experiments have been performed to validate the method. Our results show that the shear moduli of soft composites consisting of matrix and inclusions of several millimeters in cross-section dimensions with either regular or irregular geometries can be identified with good accuracy. The advantages of the SWENet over conventional SWE methods consist of using more features of the wave motion in inhomogeneous soft materials and enabling seamless integration of multi-source data in the inverse analysis. Given the advantages of the reported method, it may find applications including but not limited to mechanical characterization of artificial soft biomaterials, imaging elastic properties of nerves in vivo, and differentiating small malignant tumors from benign ones by quantitatively measuring their distinct stiffnesses.

**Keywords**: shear wave elastography; physics informed neural networks; inverse problem; tissue-mimicking phantom experiments

## 1   Introduction

Most soft materials including soft tissues have large Poisson ratios and small shear moduli. This feature determines that the velocity of a pressure wave in them is much faster than that of a shear wave. The characteristics of reflection and scattering of high frequency pressure waves (e.g., ultrasonic wave) in soft materials have been used in medical imaging, for instance ultrasound imaging (Donald et al., 1958). Whereas the slow propagation speed of shear waves can be tracked with ultrafast imaging methods and forms the basis of shear wave elastography (SWE), which has received considerable attention across different disciplines during the past two decades (Bercoff et al., 2004; Sarvazyan et al., 1998; Song et al., 2015).

SWE enables the measurement of constitutive parameters (e.g., linear elastic, viscoelastic and hyperelastic parameters) of soft tissues *in vivo* and finds broad clinical applications because the occurrence and development of diseases are usually accompanied by the variation of their mechanical properties (Cao et al., 2019; Cosgrove et al., 2013). For example, SWE has been used to stage liver fibrosis (Ferraioli et al., 2015) , detect artery stiffening (Li et al., 2022) and differentiate malignant tumors from benign ones (Asteria et al., 2008; Chang et al., 2011). Considering the complexity of the constitutive behavior of soft tissues and diverse clinical applications, different SWE methods have been developed in the literature (Bercoff et al., 2004; Chen et al., 2009; Gennisson et al., 2007; Jiang et al., 2015; Konofagou et al., 2001; Nazari and Barbone, 2018; Zheng et al., 2021). A common feature in these methods is that shear wave velocities are the main inputs of the inverse analysis to infer tissue mechanical properties. It is of notice that most soft tissues are inhomogeneous and wave dispersion and scattering may occur; in this case, an accurate measurement of wave velocities and further infer the mechanical properties are by no means trivial. Previous study demonstrated strong size effects observed in SWE of a solid cylindrical inclusion given by conventional SWE methods (Song et al., 2015), in which large errors occur when the cross-section dimension of a phantom tumor is smaller than 1 cm. To overcome this limitation, including more features of the elastic waves in inhomogeneous soft materials in the inverse analysis instead of merely wave velocities is necessary and important.

Deep learning (DL) relies on an improved artificial neural network, consisting of multiple processing layers to learn the representative features of data, and has been demonstrated to be useful for medical image analysis (Chen et al., 2022; Litjens et al., 2017). Being able to include the spatial and temporal features of wave motion in training, DL-based SWE has the potential to provide a more accurate evaluation of spatial distribution of tissue mechanical properties in comparison with conventional SWE. Indeed, efforts towards this direction have been made in recent years (Ahmed et al., 2021; Neidhardt et al., 2022; Vasconcelos et al., 2021; Wang et al., 2019; Zhang et al., 2016). Given that sufficient dataset required for training a deep neural network is difficult to achieve *in vivo*, numerical simulations and phantom experiments have been used to generate the required training dataset (Ahmed et al., 2021; Neidhardt et al., 2022; Vasconcelos et al., 2021). However, the accuracy of a DL-based SWE trained in this way largely depends on the extent to which the feature of elastic waves in target solids, e.g., in soft tissue *in vivo*, can be captured by the simulations or phantom experiments. Considering the difficulties encountered in SWE methods based on convolutional neural network (CNN) or other deep learning neural networks in achieving large dataset, in this study, we propose a SWE method based on physics-informed neural networks (SWENet). PINN encodes the governing equations of a physical problem, e.g., partial differential equations (PDE), as a part of the neural network training and have emerged as a useful tool to solve both direct and inverse problems (Raissi et al., 2019). Very recent studies have demonstrated that PINN is promising for full-wave inversion in geophysics from limited data (Rasht-Behesht et al., 2022). Unlike the data achieved in geophysics, stimulus is programmable in SWE of soft materials to generate multi-source data and desired wave fields required in a reliable inverse analysis, which can be well integrated with PINN to probe local mechanical properties of inhomogeneous soft materials as demonstrated in this study.

The paper is organized as follows. In section 2, governing equations for the wave motion in an inhomogeneous soft material are first briefly described, which are used subsequently to develop PINN-based shear wave elastography (SWENet). To demonstrate the usefulness of the proposed method, finite element analysis and tissue-mimicking phantom experiments have been performed to generate the dataset used in training the networks and infer the material parameters. Section 3 shows the results of learning hidden elastic properties of inhomogeneous soft materials from wave fields given by FEA and phantom experiments. In section 4, salient features of SWENet and speeding up SWENet by transfer learning have been discussed. Section 5 gives the concluding remarks.

## 2  Material and Methods

### 2.1  Wave motion in elastic soft materials

The wave equation that will be encoded into the deep neural network is simply described here. The equation of motion that governs the elastic wave propagation is

$$\frac{\partial \sigma_{ij}}{\partial x_j} = \rho \frac{\partial^2 u_i}{\partial t^2}, \tag{1}$$

where $\sigma_{ij}$ is the Cauchy stress tensor. $\rho$ is the mass density. $u_i$ is the displacement. $x_i$ ($i = 1,2,3$) and $t$ are spatial coordinate and time, respectively. Here, we consider isotropic incompressible materials for which the constitutive relation is given by

$$\sigma_{ij} = \mu \left( \frac{\partial u_i}{\partial x_j} + \frac{\partial u_j}{\partial x_i} \right) - p_0 \delta_{ij}, \tag{2}$$

where $\mu$ is the shear modulus; in general, it varies with spatial coordinates $\mu = \mu(x_1, x_2)$. $\delta_{ij}$ denotes Kronecker delta. $p_0$ is a Lagrange multiplier associated with the incompressible constraint, i.e.,

$$u_{k,k} = 0. \tag{3}$$

Inserting Eq. (2) into Eq. (1) gives

$$\rho \frac{\partial^2 u_i}{\partial t^2} = \mu \frac{\partial^2 u_i}{\partial x_j \partial x_j} + \frac{\partial \mu}{\partial x_j} \left( \frac{\partial u_i}{\partial x_j} + \frac{\partial u_j}{\partial x_i} \right) - \frac{\partial p_0}{\partial x_i} \tag{4}$$

Equation (4) denotes the wave equation expressed in terms of displacements. For SWE, the particle velocity $v_i = \frac{\partial u_i}{\partial t}$ is frequently used. We can take the time derivate of Eq. (4) and get wave equation expressed in terms of $v_i$. For a plane wave in $(x_1, x_2)$, the wave equations can be written as

$$\begin{cases} \mu \left( \frac{\partial^2 v_1}{\partial x_1^2} + \frac{\partial^2 v_1}{\partial x_2^2} \right) + 2 \frac{\partial \mu}{\partial x_1} \frac{\partial v_1}{\partial x_1} + \frac{\partial \mu}{\partial x_2} \left( \frac{\partial v_1}{\partial x_2} + \frac{\partial v_2}{\partial x_1} \right) - p_{,1} - \rho \frac{\partial^2 v_1}{\partial t^2} = 0 \\ \mu \left( \frac{\partial^2 v_2}{\partial x_1^2} + \frac{\partial^2 v_2}{\partial x_2^2} \right) + 2 \frac{\partial \mu}{\partial x_2} \frac{\partial v_2}{\partial x_2} + \frac{\partial \mu}{\partial x_1} \left( \frac{\partial v_1}{\partial x_2} + \frac{\partial v_2}{\partial x_1} \right) - p_{,2} - \rho \frac{\partial^2 v_2}{\partial t^2} = 0 \end{cases}, \tag{5}$$

where $p = \partial p_0 / \partial t$. And the updated incompressibility constraint reads

$$v_{1,1} + v_{2,2} = 0. \tag{6}$$

We introduce a stream function $\psi = \psi(x_1, x_2, t)$ that satisfies

$$v_1 = \frac{\partial \psi}{\partial x_2}, v_2 = -\frac{\partial \psi}{\partial x_1}, \tag{7}$$

which makes Eq. (6) automatically satisfied.

The wave motions are completely described by $\psi$ and $p$, both are functions of spatial coordinates and time. In contrast, the shear modulus $\mu$ varies with coordinates but is time independent. We therefore will use separate neural networks for $\psi$ and $p$, and $\mu$.

## 2.2 SWENet: a physics-informed deep neural network for shear wave elastography

Here we propose SWENet, which consists primarily of two deep neural networks, as shown in Fig. 1. In brief, neural network 1 (NN1) takes the spatial coordinates $(x_1, x_2)$ and time $t$ as input. Whereas neural network 2 (NN2) takes $(x_1, x_2)$ as input. The outputs of the neural networks are enforced to match the data and satisfy the physical constraints (i.e., the partial differential equations, PDE) after training. Therefore, the output of NN1 will closely approximate $\psi$ and $p$. And more importantly, the NN2 will infer the shear modulus in the region-of-interest (ROI) $\mu_{\mathrm{ROI}}$, the ultimate goal of SWE.

As shown in Fig. 1, ROI is a subdomain of the full field with an arbitrary shape; any points within ROI must have been illuminated by the traveling shear waves in order to infer $\mu_{\mathrm{ROI}}$. For simplify, we took square ROIs for all the studies. We took fully connected feed-forward architectures for NN1 and NN2, as suggested in previous studies on PINN (Raissi et al., 2019; Rasht-Behesht et al., 2022). Denote $L$ the number of layers (an input layer, an output layer, and $L-2$ hidden layers) and $N_l$ the number of neurons for the $l$-th layer ($l = 1, \ldots, L$). Taking the NN1 as an example (see Table 1), $L = 10$ and the numbers of neurons for the hidden layers are identical $N_l = 40$ ($l = 2, \ldots, 9$). A fully connected feed-forward neural network is a transformation from the input $\boldsymbol{x}$ to the output $\boldsymbol{y}$

$$\begin{cases} \widehat{\boldsymbol{x}}^{(1)} = \boldsymbol{x} \\ \widehat{\boldsymbol{x}}^{(l)} = \sigma(\boldsymbol{W}^l \widehat{\boldsymbol{x}}^{(l-1)} + \boldsymbol{b}^l), 1 < l < L, \\ \boldsymbol{y} = \boldsymbol{W}^L \widehat{\boldsymbol{x}}^{(L-1)} + \boldsymbol{b}^L \end{cases} \quad (8)$$

where $\boldsymbol{W}^l$ ($\in \mathbb{R}^{N_l \times N_{l-1}}$) and $\boldsymbol{b}^l$ ($\in \mathbb{R}^{N_l}$) denote the weights and biases for the $l$-th layer (no weights and biases for the input layer), respectively. $\sigma$ denotes the activation function. In this study we used the tanh function. It can be proved that the transformation can approximate any measurable functions (Hornik et al., 1989). The weights and biases will be identified through a training process based on the back-propagation algorithm (Rumelhart et al., 1986), which aims to minimizing a loss function $\mathcal{L}$.

In SWENet, $\mathcal{L}$ primarily consists of two parts, i.e., the data driven part and the physical informed part that encodes the wave equations into the loss function. As $\psi$ is the output of NN1, $v_1$ and $v_2$ can be derived according to Eq. (7) by the automatic differentiation of $\psi$ (see Fig. 2). The data driven part $\mathcal{L}_{\mathrm{Data}}$ is defined as

$$\mathcal{L}_{\mathrm{Data}} = |v_2 - v_2^*|_2, \quad (9)$$

where $v_2^* = v_2^*(x_1, x_2, t)$ is measured from the traveling shear waves. '$|\ \ |_2$' denotes the L2 norm. The physics informed part is $\mathcal{L}_{\mathrm{PDE}} = \mathcal{L}_{\mathrm{PDE1}} + \mathcal{L}_{\mathrm{PDE2}}$, where

$$\begin{cases} \mathcal{L}_{\mathrm{PDE1}} = \left| \mu \left( \frac{\partial^2 v_1}{\partial x_1^2} + \frac{\partial^2 v_1}{\partial x_2^2} \right) + 2 \frac{\partial \mu}{\partial x_1} \frac{\partial v_1}{\partial x_1} + \frac{\partial \mu}{\partial x_2} \left( \frac{\partial v_1}{\partial x_2} + \frac{\partial v_2}{\partial x_1} \right) - p_{,1} - \rho \frac{\partial^2 v_1}{\partial t^2} \right|_2 \\ \mathcal{L}_{\mathrm{PDE2}} = \left| \mu \left( \frac{\partial^2 v_2}{\partial x_1^2} + \frac{\partial^2 v_2}{\partial x_2^2} \right) + 2 \frac{\partial \mu}{\partial x_2} \frac{\partial v_2}{\partial x_2} + \frac{\partial \mu}{\partial x_1} \left( \frac{\partial v_1}{\partial x_2} + \frac{\partial v_2}{\partial x_1} \right) - p_{,2} - \rho \frac{\partial^2 v_2}{\partial t^2} \right|_2 \end{cases} \quad (8)$$

It is worth noting that $v_1$ has not been introduced into $\mathcal{L}_{\mathrm{Data}}$ because currently only $v_2^*$ is available in experiments. It will be straightforward to incorporate $v_1$ into $\mathcal{L}_{\mathrm{Data}}$ if $v_1^*$ is measurable. We expect improved performance of SWENet from introducing additional $v_1^*$. To balance the contributions of data and physics, we define

$$\mathcal{L} = \frac{1}{M} \lambda_{\mathrm{PDE}} \mathcal{L}_{\mathrm{PDE}} + \frac{1}{M} \lambda_{\mathrm{Data}} \mathcal{L}_{\mathrm{Data}}, \quad (9)$$

where $\lambda_{\mathrm{PDE}}$ and $\lambda_{\mathrm{Data}}$ are hyperparameters that should be optimized, especially when noisy data are involved. $M$ denotes the number of the datapoints used for training, i.e., batch size. $M = M_i + M_o$, where $M_i$ and $M_o$ denote the numbers of datapoints taken from inside and outside of the ROI, respectively. The datapoints in the batch are randomly chosen from the spatiotemporal space ($x_1 \times x_2 \times t$) and will be updated every 1,000 epochs.

Table 1 Architectures of the fully connected feed-forward neural works used in SWENet

| NN | Input variables | Output variables | Hidden layers | Neurons of hidden layers |
|---|---|---|---|---|
| NN1 | 3 | 2 | 8 | 40 |
| NN2 | 2 | 1 | 4 | 20 |

## 2.3 Finite element simulations

We performed Finite element simulations to produce simulation data that were used to validate SWENet. All the simulations were performed with the help of Abaqus/standard (Abaqus 6.14, Dassault Systèmes®). We created a two-dimensional square domain that was large enough to avoid wave reflections at the boundaries. The constitutive model for the material was incompressible neo-Hookean. For illustration, in all numerical examples and phantom experiments performed below, the shear modulus outside the ROI is supposed to be a known constant $\mu_0$ that can be measured, for example, by shear wave speed. Then the shear modulus for the full filed can be expressed as

$$\mu(x_1, x_2) = \begin{cases} \mu_{\text{ROI}}, (x_1, x_2) \in \text{ROI} \\ \mu_0, \text{otherwise} \end{cases}. \tag{12}$$

The ROI was a subdomain within the model in which shear waves would completely illuminate. The spatial distribution of the shear modulus was prescribed by defining a temperature-dependent shear modulus and then specifying the temperature field. The density $\rho$ was a constant 1000 kg/m³ throughout the model.

The simulations were performed in the time domain. To generate shear waves, in the first analysis step, we applied a vertically moving body force $\boldsymbol{f}$,

$$\boldsymbol{f}(x_1, x_2, t) = \boldsymbol{f}_0 \exp\left[-\frac{(x_1 - x_{10})^2}{2r_1^2} - \frac{(x_2 - x_{20} - v_a t)^2}{2r_2^2}\right], \tag{10}$$

where $\boldsymbol{f}_0 = [0, f_0]$ is a vector that specifies the magnitude and direction of the peak body force. We took a small value for $f_0$ to ensure a small amplitude for the shear waves. $v_a$ denotes the moving speed of the force, which was set to be $40\sqrt{\mu_0/\rho}$. The greater speed than these of the shear waves gives rise to the elastic Cherenkov effect, resulting in a Mach cone of the shear waves (Li et al., 2016). The parameters $r_1 = 0.14$mm and $r_2 = 1.1$ mm specify the spatial distribution, and the coordinate $(x_{10}, x_{20})$ specifies the location of the body force. The duration of the body force was about 0.1 ms. The moving body force simulates the acoustic radiation force (ARF) generated by an ultrasound sound beam that successively focuses on different locations along $x_2$ axis. A subsequent analysis step was used to compute the shear wave propagation. The duration of this step was 5 ms and we output the particle velocities $v_1^*$ and $v_2^*$ every 0.1 ms.

We used a uniform mesh with an element size of $\sim 0.05 \times 0.05$ mm². The element type was CPE8RH in Abaqus (plane strain, 8-node biquadratic, hybrid with linear pressure and reduced integration). Convergence of simulations results was checked by decreasing the element size and finding the change in the results is smaller than 1%. After the simulation, we down-sampled the results based on a uniform grid size of $0.05 \times 0.1$ mm², making the dataset consistent with our experiments.

## 2.4 Tissue-mimicking phantom experiments

In experiments, we performed SWE on a gelatin phantom. Following the protocol described in our previous paper (Zhang et al., 2020), we prepared a phantom that has a stiffer cylindrical inclusion with a radius of ~3 mm. Briefly, we prepared the matrix and the inclusion separately, each with different concentrations of carrageenan (weight fraction 0.1% vs. 0.2%) to tune the stiffness. Taking the preparation of the inclusion as an example, we dissolved gelatin (5%) and carrageenan (0.2%) into 80℃ water and then added cellulose (Type 40, Mreda, Beijing, China) with a weight fraction of 2% into the solution to serve as ultrasound scatters. We put the solution into a designated mold and then cured the solution in a 4℃ fridge for 2h.

Our experimental system was built on a Verasonics Vantage 64LE (Verasonics Inc., Kirkland, WA, USA) with an L9-4 transducer (central frequency 7 MHz, 128 elements, pitch 0.3 mm). Shear waves were generated by focused ultrasound beams using 32 elements (voltage ~10 V, aperture size ~10 mm, and uniform apodization). Six ultrasound beams (the duration of each beam was 0.042 ms) successively focused at 7.1 mm to 22.6 mm deep, resulting in a moving ARF. After the excitation, the system switched to the plane wave imaging mode working at a frame rate of 10 kHz to acquire the shear wave propagation. In this mode, all the 128 elements (voltage ~10 V, aperture size ~40 mm, uniform apodization) were used to transmit, but only the 64 elements at the center of the transducer were used to receive. We simply used the delay and sum beamforming for the plane wave imaging. The in-phase and quadrature (IQ) data were analyzed offline to get the particle velocity $v_2^*$ using the Loupas' estimator (Loupas et al., 1995). To improve the signal-to-noise ratio (SNR) of the data, we performed thirty successive measurements, and the average of the measurements was taken as the training data.

The training data for SWENet was obtained in the cross-section of the cylindrical inclusion. However, to estimate the elastic properties phantom, we also measured the shear wave speeds in the inclusion and the softer matrix, respectively, in the longitudinal view, i.e., shear waves propagate along the axial direction. The group velocity of the elastic waves in the inclusion is approximately equal to that of the bulk shear waves because the waveguiding effect is negligible owing to a relatively large inclusion radius; and in this case, the shear moduli of the inclusion and the matrix were calculated using $\mu = \rho v_s^2$, where $v_s$ denotes the shear wave speed.

## 3    Results

### 3.1    Learning hidden elastic heterogeneities from travelling shear waves with SWENet

We first show how SWENet infers hidden elasticity from traveling shear waves. Figure 2a shows representative snapshots ($t = 0$ ms, 1.5 ms, 3 ms, and 4.5 ms) of the simulation data used for training (denoted by $v_2^*$). The dashed square (11×11 mm²) shows the ROI, in which the shear modulus has a Gaussian shape with a radius of 1.5 mm. The shear modulus at the center of the ROI is the peak value of 6 kPa, then decreases to a low plateau of 4 kPa. The shear waves are generated on the left side of the ROI. The waves propagating to the right-hand direction illumine the inclusion, and then the planar wave front is distorted by elastic heterogeneity. The number of the datapoints inside and outside the ROI are 24531 and 56070, respectively, and the batch sizes of the NN1 and NN2 are $M_i = 8e3$ and $M_o = 1.5e4$. We updated the batches every 1000 epochs. We tried different weight ratios $\lambda_{\text{PDE}}/\lambda_{\text{Data}}$ (0.1, 1, 10) and found it has negligible effects on the results for this example. In this case, we take $\lambda_{\text{PDE}} = 1$ and $\lambda_{\text{Data}} = 0.1$.

Figure 2b depicts the evolution of each loss term ($\mathcal{L}_{\text{PDE}}$ and $\mathcal{L}_{\text{Data}}$). At the first ~2e3 epochs, $\mathcal{L}_{\text{Data}}$ decreases dramatically but $\mathcal{L}_{\text{PDE}}$ increases, due to the low contribution of $\mathcal{L}_{\text{PDE}}$ to the total loss function. After the first 2e3 epochs, $\lambda_{\text{PDE}}\mathcal{L}_{\text{PDE}} \approx \lambda_{\text{Data}}\mathcal{L}_{\text{Data}}$, so the two-term decrease simultaneously with similar speeds. The total loss reaches a plateau after about 7e5 epochs. Therefore, we stopped the training at 8e5 epochs. The total loss is reduced by more than four orders of magnitude after the training. In Fig. 2a, we show $v_2$ output by the NN1 and the comparison with the training data $v_2^*$. The relative error, defined as $|(v_2 - v_2^*)/\max(v_2^*)|$ for each snapshot, is less than 3%, indicating an excellent fit to the training data. Notably, despite the fact that we didn't use $v_1^*$ in the training as this data is not available for ultrasound SWE, the NN1 correctly retrieves $v_1$, with good accuracy (relative error less than 15%, see Fig. 2c). The ability of NN1 to retrieve $v_1$ should ascribe to the incompressible constraint (see Eq. (6)), revealing the benefits of incorporating physical information into deep neural networks.

SWENet also learns the hidden elastic heterogeneity within the ROI. As the NN2 is a surrogate model of the shear modulus in the ROI, it has been trained to enforce the wave equations satisfied. Figure 2d shows the shear modulus output by the NN2, which is in excellent agreement with the Gaussian-shaped distribution of the shear modulus ($\mu^*$) used in the simulation. To make a quantitative comparison, we also plot the transverse profiles of the shear moduli through the centers of the maps. The maximum relative error between the two curves is less than 2%.

For this simple case, we find the SWENet greatly outperforms the traditional method used in SWE, i.e., the time-of-flight (ToF) algorithm (Deffieux et al., 2011). With ToF, local shear wave speed can be estimated based on the time delay between a given distance. ToF is a robust algorithm and is widely used in SWE. However, as shown in Fig. 2d, ToF fails to get the shear wave speed near the ARF due to the lack of well-developed wave profiles. And because of the diffraction of the shear waves, the shape of the inclusion identified by ToF deviates from a circular shape, and a fake wake emerges. In fact, wave reflections and diffractions are primary factors that may lead to bias in the ToF (Deffieux et al., 2011; Zhao et al., 2011), making the inference of complex elastic heterogeneities from traveling waves extremely difficult. Next, we show the ability of SWENet to integrate multiple-source data to infer complex elastic heterogeneities.

### 3.2 Integrating multi-source data with SWENet

In the state-of-the-art SWE measurements, acoustic radiation force (ARF) that can be programmed both spatially and temporally is frequently used to induce shear waves in soft materials. While ARF enables a flexible way to introduce multiple source shear waves, it remains to be a challenge to integrate the measurements from multiple source shear waves (Tanter et al., 2008). Here we show that multi-source data given by programmed ARFs can be seamlessly integrated into SWENet to achieve more accurate results in the inverse analysis.

To demonstrate this significant advantage in SWENet, we consider an example shown in Fig. 3. The ROI (30×12mm$^2$) contains the letters "THU", in which the shear modulus is 24 kPa; otherwise, the shear modulus is 16 kPa. The complex distribution of the shear modulus will cause multiple reflections that are not able to be analyzed by ToF. For illustration, we consider two different sources of shear waves. In the first case, the ARFs were imposed between the letters "T" and "H" (Fig. 3a, first row), and in the second case, the ARFs were imposed between the letters "H" and "U" (Fig. 3a, second row). For both cases, the duration of the simulations were 3 ms. We generated the merged data by linear superpositions of the two simulations, as shown in Fig. 3a (third row). We then trained the SWENet by the single source data (L) and the merged data, respectively. We updated the batches every 5e3 epochs.

Figure 3b shows the shear moduli within the ROI inferred by SWENet. Each row shows the temporary results after the given number of epochs (i.e., 2e4, 3e4, 4e4, and 5e4). For the single source data, the letters "T" and "H" are identified; however, the letter "U" can not be inferred because of the diffractions of the shear waves induced by the strong elastic heterogeneities. With the merged data, we found a significant improvement in the performance of the SWENet; all the three letters have been clearly identified. This result clearly shows that integrating multi-source data in SWENet enables the mechanical characterization of inhomogeneous materials for which the spatial distributions of elastic moduli are complex.

### 3.3 Effect of data noise on the inferring material parameters with SWENet

We proceed to study the effect of noise on the performance of SWENet. To this end, we introduced white noises to the simulation data. To quantify, we define the signal-to-noise ratio (SNR) as $20\log_{10}(\max(|v_2^*|)/n)$, where $\max(|v_2^*|)$ is the maximum value of the signal and $n$ is the standard deviation of the noise floor. For illustration, we introduced 30 dB SNR to the simulation data, as shown in Fig. 4a. For noisy data, we find the inference in the shear modulus shows dependency on the hyperparameter $\lambda_{\text{PDE}}/\lambda_{\text{Data}}$. Figure 4b depicts $v_2$ (snapshot at $t = 3$ ms) output by the NN1 after 5e4 epochs for different weight ratios $\lambda_{\text{PDE}}/\lambda_{\text{Data}}$ ($10^1, 10^2, 10^3$, and $10^4$). The noises have been learned when the weight of data is large. However, $v_2$ becomes much smoother as the weight of the physical constraint is enhanced, suggesting that the physical constraint is an effective spatial filter. Figure 4c shows the shear modulus $\mu$ inferred by SWENet. Notably, an over large weight for the physics doesn't guarantee better performance on inference. To evaluate the accuracy of $\mu$, we calculated the root mean square error (RMSE) between $\mu$ and $\mu^*$, as shown in Fig. 4d. The plot suggests an optimal ratio $\lambda_{\text{PDE}}/\lambda_{\text{Data}} \approx 1e3$, which results in the minimal RMSE. While we expect the optimal ratio is a function of the SNR, we find for 20 dB, 30 dB, and 40 dB data, $\lambda_{\text{PDE}}/\lambda_{\text{Data}} \approx 1e3$ among $\{10^1, 10^2, 10^3, 10^4\}$ results in the best performance for SWENet. Figure 4e shows the identified shear moduli and the corresponding RMSE for data

Yin *et al.*: SWENet A PHYSICS-INFORMED NEURAL NETWORK FOR SWEwith different noise levels. Despite the success in identifying the inclusion, noise dramatically deteriorates the accuracy of SWENet, which underscores the importance of high-quality data.

### 3.4  Elastogram reconstruction for tissue-mimicking phantom with SWENet

In experiments, we applied stimuli on left and right side of the ROI in two independent measurements (see Fig. 5a), which allowed us to generate a multiple source dataset, as shown in Fig. 5b (first row). The triangles indicate the two shear wavefronts generated by the left and right stimuli, respectively. We fed the merged data to SWENet to infer the shear modulus in the ROI. In the present case, we set $\lambda_{\text{PDE}} = 1, \lambda_{\text{Data}} = 1\text{e} - 4, M_i = 8\text{e}3$, and $M_o = 8\text{e}3$. Please note a large weight for PDE was used in to suppress the noise. Therefore, we get smooth wave fields as shown in the second row of Fig. 5b. For comparison, we also trained SWENet with the dataset generated by the stimulus imposed on the right of the ROI, using the same hyperparameters. We reconstructed the shear modulus by ToF as well.

Figure 5c shows the maps of the shear moduli and the corresponding transverse profiles through the center of the inclusion. The dashed line denotes the ground truth obtained from bulk shear wave speeds, i.e., 2.3m/s in the softer matrix and 2.9m/s in the stiffer inclusion (see Methods). In both cases, we find SWENet correctly identifies the stiffer inclusion. However, as we expect, the multiple source data result in a better performance of SWENet. With the right stimulus data, SWENet performs better in identifying the right side of the inclusion, not only because of the diffraction effect (see Sec. 3.2) but also because of the attenuation of the shear waves induced by viscoelasticity of the phantom. The comparison demonstrates the feasibility of using multiple source data to get a better elastogram reconstruction in practical applications.

We find SWENet outperforms ToF in terms of suppressing artifacts. The shear modulus map reconstructed by ToF shows multiple local extremums (Fig. 5c), which are supposed to be artifacts introduced by noises. When dealing with experimental data low SNR, SWENet with wave equations encoded into enables to filter out noises, and thus suppress the non-physical artifacts. However, spatial filters that are applied for pre- and post-treatments involved in ToF (Deffieux et al., 2011) don't take physical law into consideration and thus potentially introduce artifacts usually encountered in the clinical used of ultrasound SWE.

### 4  Discussion

Our study demonstrates that SWENet enables inferring the spatial distribution of elastic properties from traveling shear waves with higher accuracy in comparison with conventional SWE methods relying on the measurement of shear wave velocities. Identification of material parameters from traveling waves is a typical inverse problem. There are several merits of SWENet in dealing with such an inverse problem. First, owing to the full-waveform inversion approach, SWENet makes full use of the data to infer the elastic properties instead of merely calculating the local shear wave speed with the time-of-flight algorithm. Sufficient information is the key to guaranteeing the uniqueness and instability of the identified solution. Indeed, a lack of information can not be remedied by any mathematical tricks in solving an inverse problem (Lanczos, 1961). Second, SWENet encodes the governing equations of wave motion in loss function, which essentially acts as a regularization term to constrain the space of admissible solutions. In contrast, conventional regularization approaches usually suffer from the drawback of lacking physical meaning. Third, compared to the supervised learning-based methods, the PINN-based SWENet doesn't require a vast dataset for training and therefore retains the flexibility to choose different ways to generate shear waves and designate the shapes of the ROI. Finally, SWENet allows us to integrate multi-source data in the inverse analysis, which is particularly useful for the elastography of a soft material for which the spatial distribution of elastic modulus is complex.

Regarding the ability to integrate multiple-source data, the examples we show in this study are based on two shear waves propagating along transverse directions. As shown in Fig. 5c, feeding SWENet with the merged data dramatically improves the inference in the transverse direction but is less helpful for the longitudinal direction. We expect the inference in the longitudinal direction will significantly benefit from additional tilt plane shear waves, which are possible by focusing the ultrasound beams along tilt paths. Although integrating multiple-source shear waves help infer complex hidden elasticities, we note there is a fundamental limitation on the spatial resolution of SWE, primarily determined by the wavelength of the shear wave. We envision the idea to integrate multiple-source data can easily be extended to integrating multiple-fidelity data obtained from different imaging modalities, such as ultrasound SWE, magnetic resonance elastography (Muthupillai et al., 1995), and optical coherence elastography(Kennedy et al., 2017), an approach to probe multiscale mechanical properties in situ.

Although SWENet outperforms traditional algorithms for SWE (e.g., ToF) in several aspects, training SWENet is time-consuming due to the large number of epochs needed. To speed up the training, we propose to train SWENet based on pre-trained neural networks, or transfer learning(Narkhede et al., 2022). For illustration, we consider the inference of a Gaussian-shaped inclusion with the well-trained SWENet used in Fig. 2. We reduced the Gaussian radius to 1 mm, increased the peak shear modulus to 8 kPa, and moved the center 0.5 mm along the left-handed direction. Then we performed simulations to get training data for this new inclusion. Figure 6a plots the loss functions for the pre-strained SWENet (blue line) and a randomly initialized SWENet (gray line). To achieve three orders of magnitude decrease in the loss function, it takes about 2e3 epochs for the pre-trained SWENet, whereas more than 2e4 epochs for the randomly initialized SWENet. Figs. 6b and c show the shear modulus inferred by the pre-trained SWENet. Since the weights and biases have been reloaded, the shear modulus output by the pre-trained SWENet is similar to that shown in Fig. 2d at the beginning of the training. Then the peak gradually moves to the left, and the radius decreases. After about 1e4 epochs, the shear modulus inferred by the pre-trained SWENet does not vary significantly and shows excellent agreement with the ground truth.

## 5 Conclusion

SWE of inhomogeneous soft materials remains a challenging issue due to the complexity of wave fields and their obscured correlation with material parameters. In the present study, we overcome this challenge by proposing a physics-informed neural networks (PINN)-based SWE (SWENet) method considering the merits in PINN in solving an inverse problem. The spatial variation of elastic properties of inhomogeneous materials has been defined in governing equations, which are encoded in PINN as loss functions. The particle velocities in vertical direction inside a local region that are measurable in practical experiments have been used to train the neural networks. The trained neural network enables the full-wave inversion and inferring the spatial distribution of elastic properties with high spatial resolution. Both finite element simulations and tissue-mimicking phantom experiments have been performed to validate the method. Moreover, the effect of data noise, a seamless integration of multi-source data in the inverse analysis and speeding up of the SWENet with transferring have been addressed, the results facilitate the use of SWENet in practice. Although tissue mimicking phantom experiments performed in this study have demonstrated the applicability of SWENet in mechanical characterization of soft materials, its use in clinics, for instance its application to imaging elastic properties of nerves *in vivo* and differentiating small malignant tumors from benign ones by quantitatively measuring their distinct stiffnesses, deserves further effort.

## 6 Acknowledgement

We gratefully acknowledge support from the National Natural Science Foundation of China (Grants Nos. 11972206 and 11921002).

## 7 Data availability

The data that support the findings of this study are available from the authors on reasonable request.

## 8 Code availability

The SWENet was implemented on a PC with 4 NVIDIA GeForce RTX 2080 Ti GPUs, using the deep learning framework Keras based on Tensorflow 2.5.0 (cuda 11.1, cuDNN 8.2.1). All the codes are available from the authors on reasonable request.

**Figures and captions**

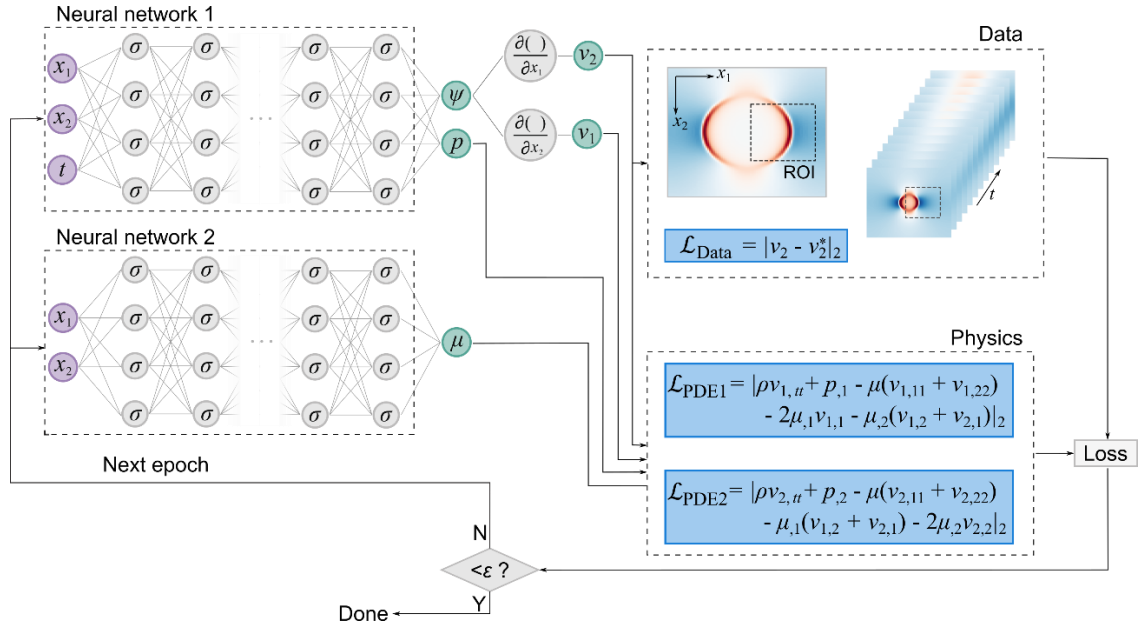

**Fig. 1 Architecture of the SWENet.** Neural network 1 (NN1) takes the spatial coordinates and time as inputs and outputs the wave stream function $\psi$ and the Lagrange multiplier $p$ for the full field. Neural network 2 (NN2) takes the spatial coordinates and outputs the shear modulus with the region-of-interest (ROI). The loss function primarily consists of the physics-informed and data-driven parts. The networks are trained to minimize the loss function.

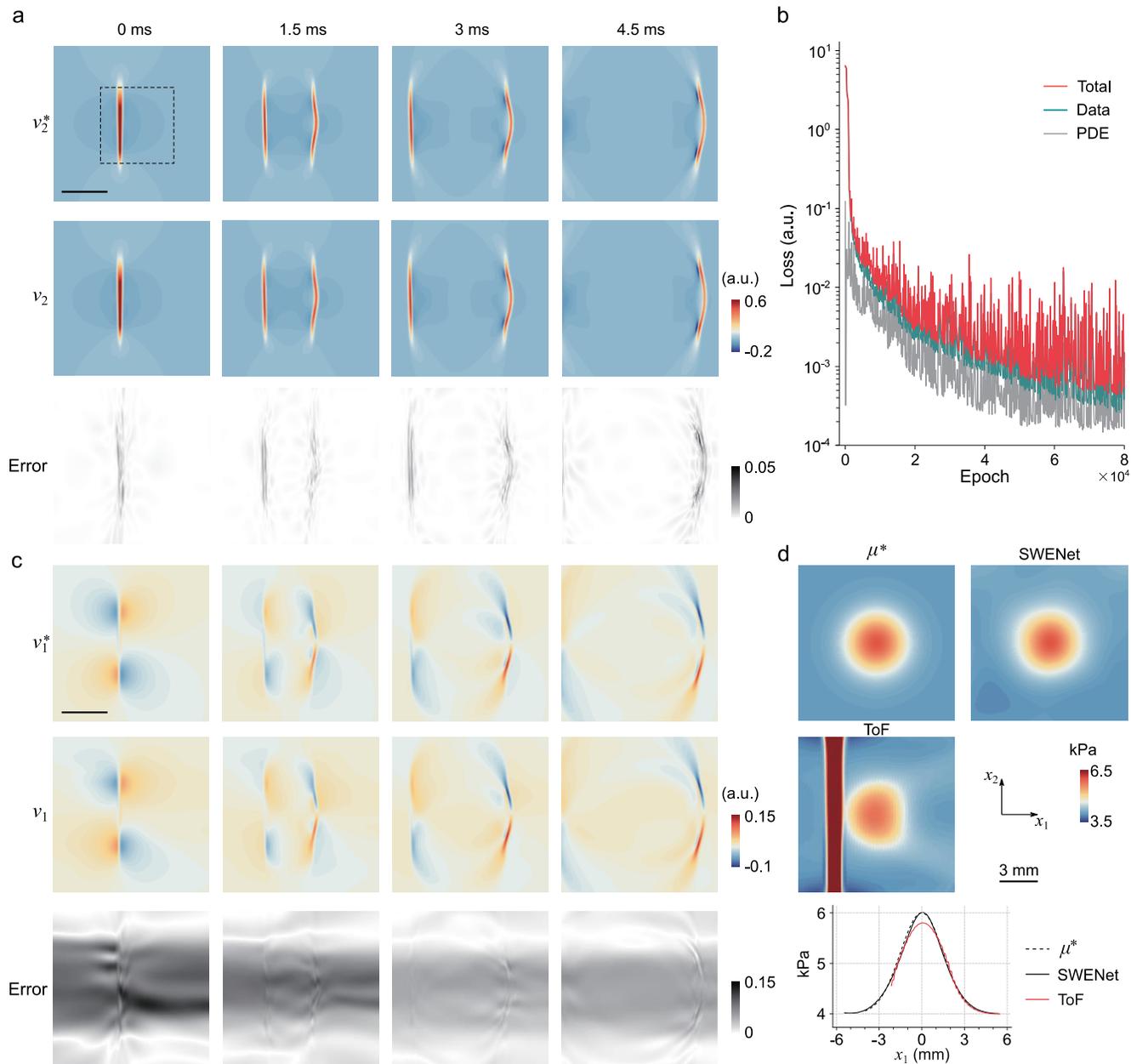

**Fig. 2 Learning hidden elasticity from traveling shear waves with the SWENet. a**, First row, the snapshots of the training data (vertical component of the particle velocity, $v_2^*$). The dashed square shows the ROI. Scalebar, 5 mm. Second row, $v_2$ output by NN1. Third row, the relative error between $v_2^*$ and $v_2$. **b**, Evolutions of the physics-informed (PDE) and data-driven terms in the loss function and their sum. **c**, First row, the snapshots of $v_1^*$ that are used to evaluate the accuracy of $v_1$ (second row) predicted by NN1. Third shows the relative error between $v_1^*$ and $v_1$. **d**, The maps of the true shear modulus ($\mu^*$), and the inference by SWENet and ToF. The plot shows the transverse profiles of the shear moduli through the centers of the maps.

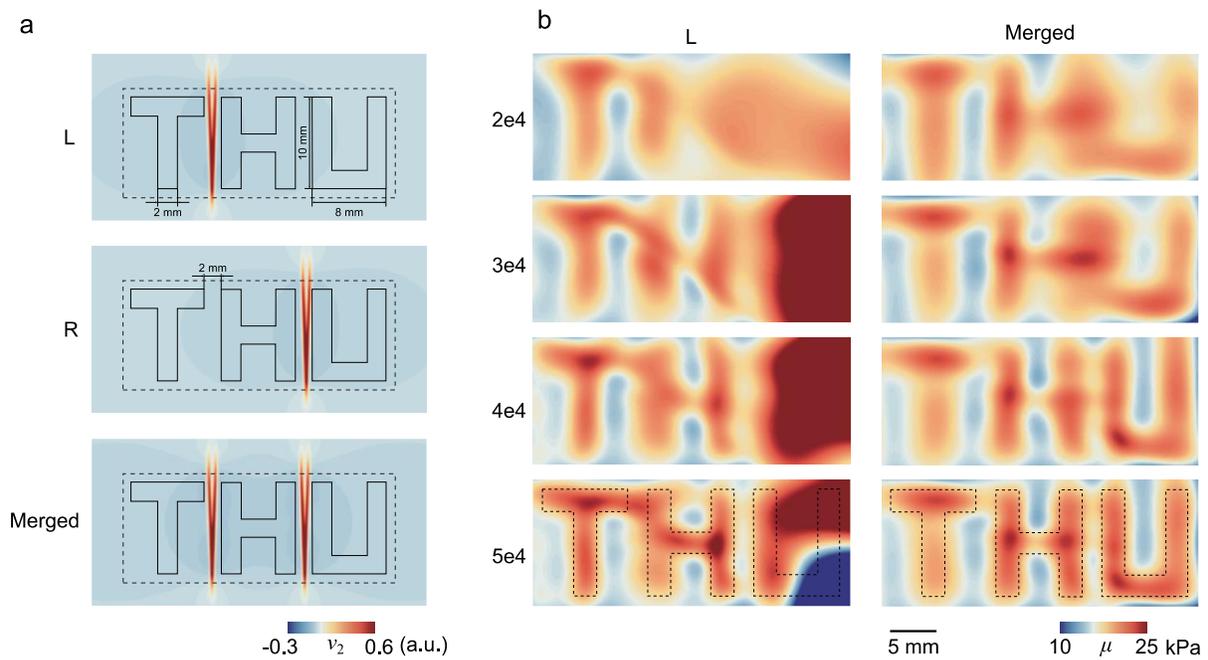

**Fig. 3 Integrating multi-source data with SWENet. a**, The first snapshot ($t = 0$ ms) of the wave motions generated by applying acoustic radiation force to left (L, first row) and right (R, second row) side of the ROI (dashed square). Third row shows the merged wave field. The shear modulus within the contour 'THU' is 25 kPa, otherwise the shear modulus is 16 kPa. **b**, Inferences of the shear moduli in the ROI based on single source (left column) and multiple-source (right column) data, respectively. Each row denotes the temporary results after the given number of epochs.

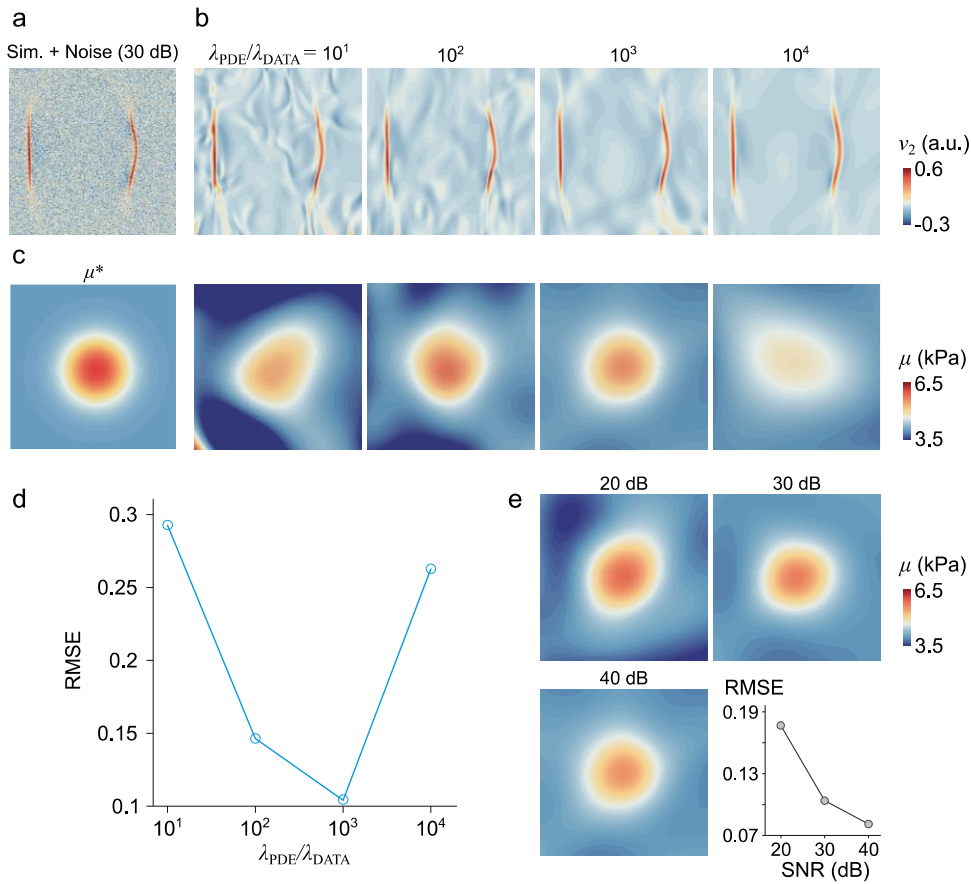

**Fig. 4 Learn hidden elasticity from noisy data with SWENet. a**, A snapshot of the simulation data ($v_2^*$) with noise added. The SNR is 30 dB. **b**, The output of NN1 ($v_2$) after 5e4 epochs when different weights for the physics ($\lambda_{\mathrm{PDE}}$) and data ($\lambda_{\mathrm{DATA}}$) have been used. From left to right, $\lambda_{\mathrm{PDE}}/\lambda_{\mathrm{DATA}} = 10^1$, $10^2$, $10^3$, and $10^4$, respectively. **c**, The inferences of the shear moduli by SWENet ($\mu^*$) and the ground truth ($\mu^*$). **d**, The RMSE of the inferences shown in c. The RMSE is minimal when $\lambda_{\mathrm{PDE}}/\lambda_{\mathrm{DATA}} = 10^3$. **e**, The shear moduli inferred by SWENet from the data with different noise levels.

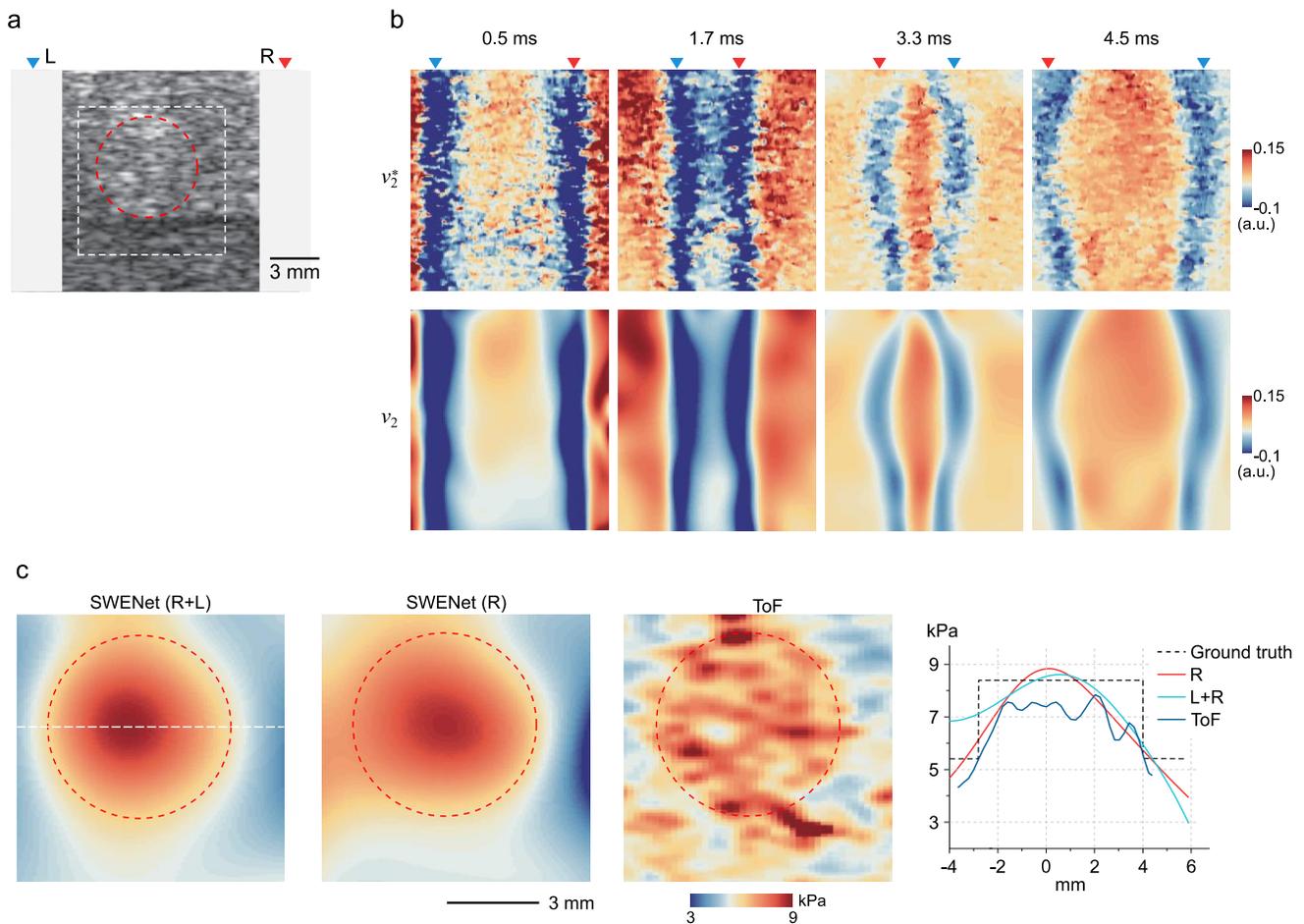

**Fig. 5 Elastogram reconstruction for a tissue-mimicking phantom based on SWENet. a**, B-mode image and photo of the phantom. The dashed square shows the ROI. The dashed circle highlights the location of the stiff cylindrical inclusion. The triangles indicate the locations of the ARF. The left (L) and right (R) stimuli are applied separately in two independent experiments. **b**, First row, snapshots of the merged experimental data (L+R). Second row, $v_2$ output from the NN1 after training on the merged data. The blue and red triangles indicate the locations of the wavefronts that correspond to L and R stimuli, respectively. **c**, The maps of the shear moduli in the ROI inferred by SWENet and ToF. The solid lines in the plot show the transverse profiles of the shear moduli through the centers of the maps. The dashed line shows the reference values of the shear moduli obtained from bulk shear wave speeds.

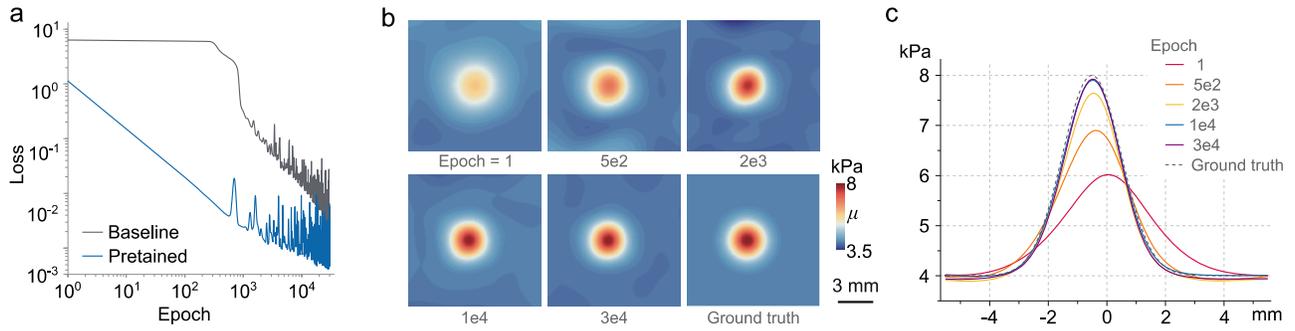

**Fig. 6 Transfer learning to speed up the elastogram reconstruction. a**, Loss functions based on randomly initialized neural networks (baseline) and pretrained neural networks. **b**, Maps of the shear modulus in the region-of-interest output at different epochs. **c**, The shear moduli along the horizontal lines passing the centers of the maps that shown in b.

19